\documentclass[12pt,a4paper]{article}
\newcommand{\ba}{\begin{eqnarray}}
\newcommand{\ea}{\end{eqnarray}}
\newcommand{\baa}{\begin{array}}
\newcommand{\eaa}{\end{array}}

\newcommand{\beq}{\begin{equation}}
\newcommand{\eeq}{\end{equation}}

\newcommand{\Cg}{{\cal C}_g}

\newcommand{\comment}[1]{}

\begin{document}

\title{\bfseries Scalar Mesons and Glueballs in a Chiral $U(3)\times U(3)$ Quark Model
with 't Hooft Interaction.}

\author{
M. Nagy{\footnote{Permanent Address:
Institute of Physics, Slovak Academy of Sciences,
842 28 Bratislava, Slovakia}} , M. K. Volkov, V. L. Yudichev\\[2mm]
\itshape Bogoliubov Laboratory of Theoretical Physics,\\
\itshape Joint Institute for Nuclear Research, 141980 Dubna, Russian Federation
}
\maketitle

\begin{abstract}
 In a $U(3)\times U(3)$ quark chiral model of the Nambu--Jona-Lasinio
(NJL) type with the 't Hooft interaction, the ground scalar isoscalar mesons
and a scalar glueball are described. The glueball (dilaton) is introduced
into the effective meson Lagrangian written in a chirally symmetric form
 on the base of scale invariance.
The singlet-octet mixing of scalar isoscalar mesons and their mixing with
the glueball are taken into account. Mass spectra of the scalar mesons and
glueball and their strong decays are described.
\vskip0.5truecm PACS: 12.39.Ki, 12.39.Mk, 13.25.-k, 14.40.-n
\end{abstract}
\thispagestyle{empty}
\newpage
\setcounter{page}{1}
\section{ Introduction}
The self-interaction of gluons, a peculiarity of QCD, gave an idea
that gluons can form bound states that can propagate particlelike
in the space. Unfortunately, because of theoretical problems,
there is still no exact answer on do these states really exist
or not. However, from recent lattice simulations
\cite{Sexton,LeeWeingarten,VaccarinoWeingarten}
one can conclude
that it is most probably that glueballs are the reality of our world.
Having assumed that glueballs exist,
one can try to construct a model to describe their interaction with other
mesons, their properties, such as,
e.g., the mass and the decay width, and to identify them with observed
resonances.

An exact microscopic description of bound gluon states cannot be done
systematically in the framework of QCD. In this situation,
QCD-motivated phenomenological models are the tool which can
help to deal with glueballs as well as with quarkonia which form the most
of observed meson states. However, using these models to describe glueballs,
we encounter many difficulties concerning, e.g., the ambiguity of the
ways of including glueballs into models and identification of
experimentally observed meson states.
This justifies the variety of points of view on this problem.

First of all, we do not know the exact mass of a glueball.
From the quenched QCD lattice simulations, Weingarten
(see e.g. \cite{Sexton,VaccarinoWeingarten}) concluded that
the lightest scalar glueball is expected around 1.7 GeV. Amsler \cite{Amsler}
considered
the state $f_0(1500)$ as a candidate for the scalar glueball.
QCD sum rules \cite{Naris_98} and K-matrix method \cite{Aniso_98}
showed that both $f_0(1500)$ and $f_0(1710)$ are mixed states with
large admixture of the glueball component.

All the bound isoscalar $q\bar q$ states are subject to mixing with
glueballs, and their  spectrum  has many interpretations made by different
authors. For instance, Palano \cite{Palano}  suggested a scenario, in which
the states $a_0(980)$, $K_0^*(1430)$, $f_0(980)$, and $f_0(1400)$ form a
nonet. The state $f_0(1500)$ is considered as the scalar glueball.
T\"ornqvist et al.~\cite{Tornqvist} looked upon the states $f_0(980)$ and
$f_0(1370)$ as manifestations of the ground and excited $s\bar s$ states,
and the state $f_0(400-1200)$ as the ground $u\bar u$ state. Eef van Beveren
et al.~\cite{Beveren} considered the states  $f_0(400-1200)$ and $f_0(1370)$ as
ground $u\bar u$ states,
and the states $f_0(980)$ and $f_0(1500)$  as ground $s\bar s$ states.
Two states for each $q\bar q$ system occur due to pole doubling, which
takes place for scalar mesons in their model.
Shakin et al.~\cite{Shakin} obtained from a nonlocal confinement model that
the $f_0(980)$ resonance is the ground $u\bar u$ state and $f_0(1370)$
is the ground $s\bar s$ state. The state $f_0(1500)$ is considered as a radial
excitation of $f_0(980)$. They believe the mass of  scalar glueball to be
1770 MeV.

In our recent papers \cite{Volk_991,Volk_992,Volk_993},
following the methods given in Refs.~\cite{Volk_82,Volk_86,Weiss,YaF},
we showed that all experimentally observed scalar meson states with
masses in the interval from 0.4 GeV till 1.7 GeV can be interpreted
as members of two scalar meson nonets ---
the ground meson nonet and its first
radial excitation. We considered all scalar mesons as $q{\bar q}$
states and took into account the singlet-octet mixing  caused by
the 't Hooft interaction. However, we did not describe the scalar
state $f_0( 1500)$.
We supposed that this state contained
a significant component of the scalar
glueball (see \cite{Naris_98,Aniso_98}).

In this work, we introduce the glueball
field into our effective Lagrangian as a dilaton and  describe
it and its mixing with the  scalar isoscalar $q\bar q$ states.
We consider only ground $q\bar q$ states.
As a result, we describe three scalar mesons
$f_0(400-1200)$, $f_0(980)$, and $f_0(1500)$ (or $f_0(1710)$). Let us note that the model
we present here is just a tentative one. In this model, we probe
one of the possible ways of including the scalar glueball into a model
that would describe the glueball state and quarkonia simultaneously.

In various works  (see e.g.
\cite{Kusa_93,Andr_86,Jami_96,Elli_84}), the authors
introduced the glueball into meson  Lagrangians using
the principle of scale invariance and the dilaton model.
For example,  $SU(2) \times SU(2)$ models were
considered \cite{Kusa_93,Andr_86,Elli_84} and the
$U(3)\times U(3)$ model, however, without the 't Hooft interaction
was investigated in \cite{Jami_96}.
Here  we  follow these works and introduce
the scalar glueball state into the Lagrangian constructed in
\cite{Volk_82,Volk_86,Cimen_99}.

In this paper, we introduce the dilaton field into the phenomenological
meson Lagrangian obtained from the effective quark interaction of
the NJL type after bosonization. We take the meson Lagrangian in
a chirally symmetric form (before the spontaneous breaking of chiral
symmetry), where only  quadratic meson terms are of scale-noninvariant
form, therefore, we introduce the dilaton field only into these (massive)
terms. To this Lagrangian, we add the potential describing the dilaton
self-interaction in the form given in
 Ref.~\cite{OPE} and dilaton kinetic term.

After introducing the glueball into our effective Lagrangian we
describe the mixing of three scalar mesons: two scalar isoscalar
$(q{\bar q})$ states and the glueball. As a result, we obtain the mass
spectrum of these scalar meson states
and also describe their main strong decays.

We emphasize once more that we do not claim this model to give a quantitative
description of experimental data. We do not take into account the states $f_0(1370)$
and one of the states $f_0(1500)$, $f_0(1710)$
which could not be neglected if the real meson spectrum
were our goal.
We intend to obtain rather qualitative estimates
that will allow us to choose a more reasonable way of including the scalar
glueball field into the effective meson Lagrangian.

In Section 2, we discuss an effective meson Lagrangian
with the 't Hooft interaction. In Section 3,  we
introduce a dilaton field into the Lagrangian.
In Section 4, we derive  gap equations for
the masses of $u(d)$ and $s$ quarks. We also deduce a set of
equations allowing us to bind two of the  three new parameters connected with
the dilaton field.
In Section 5, we obtain  mass formulae for  scalar
mesons and the dilaton and describe the singlet-octet mixing
among quarkonia and the
mixing of the dilaton with two isosinglet  scalar
mesons.  Here we fix the new model
parameters and  give their numerical estimates as well as the estimates for
the masses of scalar mesons and the glueball.
In Section 6, we calculate the widths of main strong decays of the
scalar mesons and the glueball. In the Conclusion,  we   discuss
obtained results.

\section{Chiral effective Lagrangian with  't Hooft interaction}

A $U(3)\times U(3)$ chiral Lagrangian with
the 't Hooft interaction was investigated in paper \cite{Cimen_99}.
It consists of three terms as shown in formula (\ref{Ldet}).
The first term represents
the free  quark Lagrangian, the second  is composed of
four-quark vertices  as in the NJL model, and the last one describes the
six-quark 't Hooft interaction that is necessary to solve
the $U_A(1)$ problem.
\ba
L& =& {\bar q}(i{\hat \partial} - m^0)q + {G\over 2}\sum_{a=0}^8 [({\bar q}
{\lambda}_a q)^2 +({\bar q}i{\gamma}_5{\lambda}_a q)^2] -\nonumber\\
&&- K \left\{ {\det}[{\bar q}(1+\gamma_5)q]+{\det}[{\bar q}(1-\gamma_5)q]
\right\}.
\label{Ldet}
\ea
Here $\lambda_a\; (a=1,...,8)$ are the Gell-Mann matrices and
$\lambda_0 = {\sqrt{2/ 3}}${\bf 1}, with {\bf 1} being the unit matrix;
$m^0$ is a current quark mass matrix with diagonal elements $m^0_u$,
$m^0_d$, $m^0_s$ $(m^0_u \approx m^0_d)$.

The standard treatment for local quark models consists in replacing  the
four-quark vertices with bosonic fields (bosonization).
The final effective
bosonic Lagrangian appears as a result of the calculation of the quark
determinant. First of all, using the method described in
\cite{Cimen_99,Vogl_91,Kleva_92}, we go from Lagrangian (\ref{Ldet})
to a Lagrangian which contains only four-quark vertices\footnote{
In addition to the one-loop corrections of constants of four-quark vertices,
we allow for two-loop contributions from the 't Hooft interaction that
modify the current quark masses. Thus, we avoid the problem of double counting
of the 't Hoot contribution in gap equations which was encountered by
the author in \cite{Kleva_92}.
}
\ba
&&L = {\bar q}(i{\hat \partial} -
\overline{m }^0)q + {1\over 2}\sum_{a,b=1}^9[G_{ab}^{(-)} ({\bar q}{\tau}_a
q)({\bar q}{\tau}_b q) + G_{ab}^{(+)}({\bar q}i{\gamma}_5{\tau}_a q)({\bar
q}i{\gamma}_5{\tau}_b q)], \label{LGus}
\ea
where
\ba &&{\tau}_a={\lambda}_a
~~~ (a=1,...,7),~~~\tau_8  = ({\sqrt 2} \lambda_0 + \lambda_8)/{\sqrt
3},\nonumber\\
&&\tau_9 = (-\lambda_0 + {\sqrt 2}\lambda_8)/{\sqrt 3},
\nonumber \\
&&G_{11}^{(\pm)}=G_{22}^{(\pm)}=G_{33}^{(\pm)}= G \pm
4Km_sI^\Lambda_1(m_s), \nonumber \\
&&G_{44}^{(\pm)}=G_{55}^{(\pm)}=G_{66}^{(\pm)}=G_{77}^{(\pm)}= G \pm
4Km_uI^\Lambda_1(m_u), \nonumber \\ &&G_{88}^{(\pm)}= G \mp 4Km_sI^\Lambda_1(m_s),\quad
G_{99}^{(\pm)}= G,\nonumber\\ &&G_{89}^{(\pm)}=G_{98}^{(\pm)}= \pm 4{\sqrt
2}Km_uI^\Lambda_1(m_u)\nonumber\\ &&G_{ab}=0\quad (a\not=b; \quad
a,b=1,\dots,7),
\label{DefG}
\ea
\ba
    \overline{m}^0_u&=&m^0_u-32 K m_u m_s
I^{\Lambda}_1(m_u)I^{\Lambda}_1(m_s) \label{twoloopcorrect1},\\
 \overline{m}^0_s&=&m^0_s-32 K m_u^2
I^{\Lambda}_1(m_u)^2\label{twoloopcorrect2}.
\ea
Here $m_u$ and $m_s$ are  constituent quark masses and
the integrals
\ba
I^{\Lambda}_n(m_a)={N_c\over (2\pi)^4}\int d^4_e k {\theta (\Lambda^2 -k^2)
\over (k^2 + m^2_a)^n}
\label{DefI}
\ea
are calculated (for simplicity) in the Euclidean metric  and
regularized  by a simple $O(4)$-symmetric
ultra-violet cutoff $\Lambda$. The mass $m_a$ can be $m_u$ or $m_s$.


Now it is necessary
to  bosonize  Lagrangian (\ref{LGus}).
After bosonization  and renormalization of the meson fields, we obtain
\cite{Volk_82,Volk_86,Cimen_99}
\ba
L(\sigma,\phi) =
-\sum_{a,b=1}^9{g_ag_b\over 2}\left(
\sigma_a(G^{(-)})^{-1}_{ab}\sigma_b + Z\phi_a(G^{(+)})^{-1}_{ab}\phi_b
\right)  \nonumber \\
-i~{\rm Tr}\ln \left\{1 + {1\over i{\hat \partial} - m}
\sum_{a=1}^9g_a\tau_a(\sigma_a + i\gamma_5 {\sqrt Z}\phi_a) \right\},
\label{Lbar}
\ea
where $(G^{(\pm)})^{-1}$ is the inverse of $G^{(\pm)}$, and
$\phi_a$ and $\sigma_a$ are the pseudoscalar and scalar fields,
respectively.
\ba
&& g_1^2=g_2^2=g_3^2=g_8^2=g_u^2=[4I^\Lambda_2(m_u)]^{-1},\nonumber\\
&&g_4^2=g_5^2=g_6^2=g_7^2=[4I^\Lambda_2(m_u,m_s)]^{-1}, \nonumber \\
&&\quad g_9^2=g_s^2=[4I^\Lambda_2(m_s)]^{-1}, \nonumber\\
&&I^\Lambda_2(m_u,m_s)=
{N_c\over (2\pi)^4}\int d^4_e k {\theta (\Lambda^2 -k^2)
\over (k^2 + m^2_u)(k^2 + m^2_s)}=\nonumber\\
&&={3\over (4\pi)^2(m_s^2-m_u^2)}\left[m_s^2\ln\left({\Lambda^2
\over m_s^2}+1 \right) - m_u^2\ln\left({\Lambda^2\over m_u^2}+1
\right) \right],
\label{ga_0}\\
&& Z=\left(1-\frac{6m_u}{M_{A_1}^2}\right)^{-1}\approx 1.44 ,
\ea
where $M_{A_1}$ is the mass of the axial-vector meson.

 In the one-loop approximation the meson Lagrangian looks as
follows\footnote{
The expression under Tr in (\ref{Lagr14}) is written
formally for the reason of being concise and is not ready for practical use. One will
find which $g_a$ should be taken, considering the corresponding
quark-loop diagrams.}
\begin{eqnarray}
 &&L(\sigma,\phi)=
 {1\over 2}\sum_{a = 1}^9[({\partial_{\mu}}\sigma_{a})^2+
 ({\partial_{\mu}}\phi_{a})^2]+\nonumber \\
&&+
 \left[(m_u-m_u^0) (G^{(-)})^{-1}_{88}-\frac{m_s-m_s^0}{\sqrt{2}} (G^{(-)})^{-1}_{98}-8m_uI^\Lambda_1(m_u)
 \right]g_8\sigma_8- \nonumber \\
&&+
 \left[(m_u-\bar{m}_u^0) (G^{(-)})^{-1}_{89}-\frac{m_s-\bar{m}_s^0}{\sqrt{2}} (G^{(-)})^{-1}_{99}+
8\frac{m_s}{\sqrt{2}}I^\Lambda_1(m_s) \right]g_9\sigma_9- \nonumber \\
&& -{1\over 2}\sum_{a,b=1}^9 g_ag_b\left[\sigma_a(G^{(-)})^{-1}_{ab}\sigma_b+
 Z\phi_a(G^{(+)})^{-1}_{ab}\phi_b \right]+\nonumber\\
&&{1\over 4}{\rm Tr}\Biggl\{g^2\Biggl[8I^\Lambda_1(m)(\sigma^2+Z\phi^2)
 - \nonumber \\
&&-\left(\sigma^2 - {\{M,\sigma\}_+ \over g} + Z\phi^2\right)^2
+\left[\left(\sigma -{M\over g}\right), \phi\right]_-^2
\Biggr]  \Biggr\},\label{Lagr14}\\
&&C_{a}=0 \quad (a=1,\dots,7),\quad C_8=1,\quad C_9=-\frac{1}{\sqrt{2}}.
\end{eqnarray}
It is useful to notice that Lagrangian (\ref{Lagr14}) can be given
a  chirally symmetric form. For this purpose, we introduce
fields $\sigma' = \sigma -m g^{-1}$ that have nonzero expectation
values, $<\sigma'>_0\ne 0$
\cite{Volk_86}. Here
\beq
g=g_8\tau_8-\frac{g_9\tau_9}{\sqrt{2}};\qquad m=m_u\tau_8+\frac{m_s\tau_9}{2}.
\eeq
This is the form of the effective Lagrangian in the case of unbroken
chiral symmetry:
\begin{eqnarray}
&&L(\sigma',\phi)
= \frac12\sum_{a = 1}^9[({\partial_{\mu}}{\sigma'}_{a})^2+
({\partial_{\mu}}\phi_{a})^2] \nonumber \\
&&-{1\over 2}\sum_{a,b=1}^9\left[\left(g_a\sigma'_a + \mu_a
 \right)(G^{(-)})^{-1}_{ab}
\left(g_b\sigma'_b + \mu_b
\right)+ Zg_ag_b\phi_a (G^{(+)})^{-1}_{ab}\phi_b \right]+ \nonumber \\
&&+{1\over 4}{\rm Tr}\left\{ 8g^2(I^\Lambda_1(m)+m^2I^\Lambda_2(m))
(\sigma'^2+Z\phi^2) - g^2[(\sigma'^2 + Z\phi^2)^2 -
Z[\sigma',\phi]^2_-] \right\},\label{Lagr12}\\
&& \mu_a=0\quad (a=1,\dots,7),\quad  \mu_8=\bar{m}_u^0, \quad
\mu_9=-\frac{\bar{m}_s^0}{\sqrt{2}},
\end{eqnarray}
where $\sigma' =\sum_{a=1}^9\sigma_{a}\tau^{a}$ and
$\phi = \sum_{a=1}^9\phi_{a}\tau^{a}$. From (\ref{Lagr12}),
one can see that, in the one-loop approximation,
our Lagrangian  does not violate chiral symmetry if $m^0=0$, and the 't Hooft
interaction is switched off.
Let us require the chiral and scale invariance both to be properties
of the effective Lagrangian. We remind that the
QCD Lagrangian satisfies this requirement  in the chiral limit ($m^0=0$).
The scale invariance of the effective meson Lagrangian is
restored by means of a dilaton field \cite{Kusa_93,Andr_86,Jami_96}
introduced into the Lagrangian so that it provides
 a proper dimension for each Lagrangian term.
The dilaton field is introduced only into the mass terms.
 We also introduce the dilaton potential intended to
reproduce QCD scale anomaly \cite{Elli_84,Colli_77,Volk_93}.

\section{Nambu--Jona-Lasinio model with dilaton}

Following earlier works devoted to dilatons, we introduce a color singlet
dilaton field $\chi$ that experiences a potential
\ba
V({\chi})=B\left({\chi\over {\chi}_0} \right)^4\left[ \ln \left({\chi\over
{\chi}_0} \right)^4 -1 \right] . \label{chi}
\ea
 It has a
minimum at $\chi = \chi_0$, and the parameter $B$ represents the vacuum
energy, when there are no quarks. The curvature of the potential at its
minimum determines the bare glueball mass
\ba m_g = {4\sqrt{B}\over {\chi}_0}.
\label{Defm_g}
\ea
To introduce the dilaton field into the effective
meson Lagrangian (\ref{Lagr12}), we use the following principle. Insofar
as the QCD Lagrangian, with the current quark masses equal to zero, is chiral and
scale invariant, we suppose that our effective meson Lagrangian, motivated by
QCD, has also to be chiral and scale invariant in the case when the current
quark masses and the 't Hooft interaction are
 equal to zero. Then, instead of Lagrangian (\ref{Lagr12}), we obtain
\ba
&&L(\sigma',\phi,\chi)= {1\over 2}\sum_{a = 1}^9
[({\partial_{\mu}}{\sigma'}_{a})^2+
({\partial_{\mu}}\phi_{a})^2]
 \nonumber \\
&&-\biggl\{{1\over 2}\sum_{a,b=1}^9\left[\left(g_a\sigma'_a + \mu_a
\right)(G^{(-)})^{-1}_{ab}
\left(g_b\sigma'_b + \mu_b \right)+
Zg_ag_b\phi_a (G^{(+)})^{-1}_{ab}\phi_b \right] \nonumber \\
&& -{1\over 4}{\rm Tr}\left\{ 8g^2(I^\Lambda_1(m)+m^2I^\Lambda_2(m))
(\sigma'^2+Z\phi^2)\right\}\biggr\} \left( {\chi\over \chi_c} \right)^2
 \nonumber \\
&&-{1\over 4}{\rm Tr}\left\{ g^2[(\sigma'^2 + Z\phi^2)^2 -
Z[\sigma',\phi]^2_-] \right\} + {1\over 2}({\partial}_{\mu}\chi)^2
-V(\chi) ,
\label{Lagr13}
\ea
where $\chi_c$ is the vacuum expectation value of dilaton fields
$\chi = {\bar \chi} + \chi_c$, $<\chi>_0 = \chi_c$
and $<{\bar \chi}>_0=0$.\footnote{
Since the mass of current quark explicitly breaks the scale invariance
of the model, there is no need to make these terms scale invariant,
using the dilaton fields.}

By rewriting this Lagrangian in terms of quantum fields $\sigma$
and ${\bar \chi}$ with vacuum expectations equal zero, we finally obtain
\ba
L(\sigma,\phi,{\bar \chi}) = L(\sigma,\phi) +
{\Delta}L(\sigma,\phi,{\bar \chi}) ,
\label{LagrDelta}
\ea
where $L(\sigma,\phi)$ equals Lagrangian (\ref{Lagr14}), and
${\Delta}L(\sigma,\phi,{\bar \chi})$ has the form%
\ba
&&{\Delta}L(\sigma,\phi,{\bar \chi}) =
{1\over 2}({\partial}_{\mu}{\bar\chi})^2
-V'({\bar \chi} + \chi_c ) + {{\bar\chi}\over \chi_c}\left(2 +
{{\bar\chi}\over \chi_c} \right) \nonumber \\
&&\times
 \left\{
-8g_8m_u^3I^\Lambda_2(m_u)\sigma_8+
8g_9\frac{m_s^3I^\Lambda_2(m_s)}{\sqrt{2}}\sigma_9\right.\nonumber\\
&&\left.-{1\over 2}\sum_{a,b=1}^9 g_ag_b\left[\sigma_a(G^{(-)})^{-1}_{ab}\sigma_b
+ Z\phi_a(G^{(+)})^{-1}_{ab}\phi_b \right]
\right.\nonumber \\
&&
\left.
+4\sum_{a=1}^9 g_{a}^2 {\cal J}_a
(\sigma^2_{a}+Z\phi^2_{a}) \right\}, \label{Lagrchi2}
\ea
where
\ba
&&{\cal J}_a= I^\Lambda_1(m_{u})
+m^2_{u}I^\Lambda_2(m_{u}),\quad (a=1,2,3,8),\nonumber\\
&&{\cal J}_a=
\frac12\left(I^\Lambda_1(m_{u})+I^\Lambda_1(m_{s})
+\frac{(m_u+m_s)^2}{4}I^\Lambda_2(m_u,m_s)\right),\quad (a=4,5,6,7),\nonumber\\
&&{\cal J}_9=I^\Lambda_1(m_s)+m_s^2 I^\Lambda_2(m_s),
\ea
and  we used gap equations (\ref{Gapeqs}) in the terms linear over $\sigma_a$.
 The potential $V'(\chi) = V(\chi)$ + $(\chi
/\chi_c)^2A$, where
\ba
A&=& {1\over 2}\sum_{a,b=1}^8
\langle\sigma_a\rangle(G^{(-)})^{-1}_{ab}\langle\sigma_b\rangle
 -4 {\cal J}_8m_u^2-2{\cal J}_9m_s^2.
\label{A}
\ea
From this Lagrangian, one can obtain the system of equations determining
constituent quark masses (gap equations) and dilaton potential parameters $\chi_0$,
$\chi_c$ and $B$.

\section{Equations for the quark masses (gap equations) and
dilaton potential parameters $\chi_0$, $\chi_c$ and $B$}

The conditions for linear terms being absent in our Lagrangian
\ba
&&{{\delta}L\over {\delta}\sigma_8}\biggr\vert_{\phi,\sigma,{\bar\chi} = 0}=0
,\quad
{{\delta}L\over {\delta}\sigma_9}\biggr\vert_{\phi,\sigma,{\bar\chi} = 0}=0
,\quad
{{\delta}L\over {\delta}\chi}\biggr\vert_{\phi,\sigma,{\bar\chi} = 0}=0
\label{var}
\ea
lead us to the following equations
\ba
(m_u-\bar{m}_u^0)(G^{(-)})^{-1}_{88} - {{m_s-\bar{m}_s^0}\over \sqrt2}(G^{(-)})^{-1}_{89} -
8m_uI^\Lambda_1(m_u)& =& 0 , \label{gapeqbegin} \\
(m_s-\bar{m}_s^0)(G^{(-)})^{-1}_{99} - {\sqrt2}(m_u-\bar{m}_u^0)(G^{(-)})^{-1}_{98} -
8 m_sI^\Lambda_1(m_s) &=& 0 ,\label{gapeq2} \\
-4B\left({\chi_c\over {\chi}_0} \right)^3{1\over \chi_0}
\ln \left({\chi_c\over {\chi}_0} \right)^4 - {2A\over \chi_c} &=& 0.
\label{Gapeqs}
\ea
An additional equation follows from the relation
between the divergence of dilaton current $S_{\mu}$ and the gluon
condensate
\ba
\langle{\partial}_{\mu}S^{\mu}\rangle_0 &=& \left(\chi{{\partial}V(\sigma',\chi)
\over {\partial}\chi} + \sum_{a=8}^9{\sigma'_a}{{\partial}V(\sigma',\chi)
\over {\partial}\sigma'_a} - 4V(\sigma',\chi)\right)
\Biggr\vert_{\begin{array}{l}\scriptstyle \chi = \chi_c\, \\[-2mm]
\scriptstyle\sigma'_a = - {\tilde \mu_a}/g_a
\end{array}}  \nonumber \\[1mm]
&=& \Cg -2 m_u^0\langle\bar{u}u\rangle_0-m_s^0\langle\bar{s}s\rangle_0,\\
 \Cg&=&\left({11\over 24}N_c - {1\over 12}N_f \right)
\left\langle{\alpha\over \pi}G^2_{\mu\nu}\right\rangle_0,
\label{gluoncon}
\ea
where $V(\sigma',\chi)$ is the potential corresponding to
Lagrangian (\ref{Lagr13}) at $\phi = 0$, and
\ba
\Cg-2 m_u^0\langle\bar{u}u\rangle_0-m_s^0\langle\bar{s}s\rangle_0
&=& 4B\left({\chi_c\over \chi_0}\right)^4+\sum_{a,b=8}^9({\tilde \mu_a}- \mu_a)
\left(G^{(-)}\right)^{-1}_{ab}\mu_b
\label{Bchi4X}
\ea
where
\beq
{\tilde \mu}_8 = m_u,\quad {\tilde \mu}_9 = -{m_s\over {\sqrt 2}}.
\eeq
The terms proportional to current
quark masses on the
right hand side of (\ref{Bchi4X}) cancel the quark condansate contribution
on the left hand side, therefore we have
\ba
\Cg &=& 4B\left({\chi_c\over \chi_0}\right)^4.
\label{Bchi4}
\ea
Using
(\ref{twoloopcorrect1}) and (\ref{twoloopcorrect2}), one can rewrite
the gap equations (\ref{gapeqbegin}) and (\ref{gapeq2})  in
a well-known form \cite{Kleva_92}
\ba
m_u^0&=&m_u-8 G m_u I_1^{\Lambda}(m_u)-32K m_u m_s I_1^\Lambda(m_u)I_1^\Lambda(m_s),\\
m_s^0&=&m_s-8 G m_s I_1^{\Lambda}(m_s)-32K(m_u I_1^\Lambda(m_u))^2.
\ea

To define all three parameters of the dilaton potential $(\chi_0,\chi_c,B)$,
we have
to use, in addition to equations (\ref{gapeqbegin})--(\ref{gluoncon}),
the equation for the bare glueball mass to be be given in the next Section.

\section{Mass formulae for scalar isoscalar mesons and glueball}

The free part of the Lagrangian (\ref{LagrDelta}) has the form
\ba
L^{(2)}(\sigma,\phi,{\bar\chi}) &=& -{1\over 2}g_8^2[(G^{(-)})^{-1}_{88}
-8I^\Lambda_1(m_u) + 4m^2_u]{\sigma}^2_8  \nonumber \\
&-&{1\over 2}g_9^2[(G^{(-)})^{-1}_{99} -8I^\Lambda_1(m_s) + 4m^2_s]{\sigma}^2_9
  \nonumber \\
&-&g_8g_9(G^{(-)})^{-1}_{89}{\sigma}_8{\sigma}_9
-8(\Cg-A)\left(\frac{\bar\chi}{\chi_c}\right)^2  \nonumber \\
&-&{16\over
\chi_c}\left[g_8m_u^3I^\Lambda_2(m_u) \sigma_8 -
g_9\frac{m_s^3I^\Lambda_2(m_s)}{\sqrt{2}}\sigma_9 \right]{\bar\chi}. \label{BAchi}
\ea
The dilaton and its interaction with quarkonia does not change the model parameters
$m_u$, $m_s$, $\Lambda$, $G$, and $K$ fixed in our earlier paper
\cite{Cimen_99}
\ba
&&m_u=280\;\mbox{MeV},\;m_s=420\;\mbox{MeV},\;\Lambda =1.25\;\mbox{
GeV},\nonumber\\
&&G=4.38\;\mbox{GeV}^{-2},\;K=11.2\;\mbox{GeV}^{-5}. \label{paramet}
\ea
After the dilaton field is introduced into our model, there appear
three new parameters: $\chi$, $\chi_c$, and $B$.
To determine these parameters, we use two equations
(\ref{Gapeqs}) and (\ref{gluoncon}) and the bare glueball mass
\beq
m_g^2=4 \frac{(\Cg-A)}{\chi_c^2}.
\eeq
We adjust it
so that, in the output, the mass of the heaviest meson would be
1500 MeV  or 1710 MeV and thereby fix $\chi_c$.
For the glueball condensate, we use the value $(390 MeV)^4$ \cite{Narison96}.
The result of our fit is presented in Table~\ref{T:spectr}
where we show the spectrum of three physical scalar isoscalar states $\sigma_I$,
$\sigma_{II}$ and $\sigma_{III}$.
\begin{table}
\centering
\caption{The masses (in MeV) of physical scalar meson states $\sigma_I$,
$\sigma_{II}$, $\sigma_{III}$ and parameter $\chi_c$
for two cases: 1) $M_{\sigma_{III}}=1500$ MeV and
2) $M_{\sigma_{III}}=1710$ MeV.}
\label{T:spectr}
\begin{tabular}{||c|c|c|c|c|c|c|c||}
    \hline
    & $\sigma_I$ & $\sigma_{II}$ & $\sigma_{III}$ & $\chi_c$ & $\chi_0$ &$B, [\mbox{GeV}^4]$& $m_g$\\ \hline
  I & 516 & 1027 & 1500 & 190 & 165 & 0.007 & 1423\\
  II & 518 & 1042 & 1710 & 198 & 172 & 0.007 & 1640\\ \hline
\end{tabular}
\end{table}
The parameters $\chi_0$ and $B$ are fixed by the gluon condensate and
constituent quark masses
\beq
\chi_0=\chi_c \exp \left({A \over 2\Cg}\right),
\eeq
\beq
B=\frac{\Cg}{4}\exp \left(-{2A \over \Cg}\right).
\eeq

It is worth noting that the mixing of the glueball with quarkonia
shifts the quarkonia mass spectrum.
They become lighter, whereas the glueball becomes heavier. This is good for
our model because, in previous paper \cite{Cimen_99}, the mass of
$\sigma_{II}$ associated with $f_0(980)$ was exaggerated.

\section{Decay widths}

Once all parameters are fixed, we estimate decay widths for the major strong decay
modes of the scalar mesons: $\sigma_a\to\pi\pi$ and $\sigma_a\to KK$.
We neglect decays into $\eta\eta$ and $\eta\eta'$ as they are small.
The results are displayed below.
The state $\sigma_{I}$ that we identify with $f_0(400-1200)$ decays mostly
into a pair of pions, and this process determines the width of $\sigma_I$:
\beq
\Gamma_{\sigma_I\to\pi\pi}\approx 600 \mbox{ MeV}.
\eeq
The decay of the state $\sigma_{II}$ that we identify with $f_0(980)$
into pions is  noticeably enhanced by the glueball
component because of mixing with the $s\bar s$ quarkonium. We obtain
\beq
\Gamma_{\sigma_{II}\to\pi\pi}=140 \mbox{ MeV}
\eeq
if $\sigma_{III}\equiv f_0(1500)$ and
\beq
\Gamma_{\sigma_{II}\to\pi\pi}=120 \mbox{ MeV}
\eeq
if $\sigma_{III}\equiv f_0(1710)$.
From experiment,
we know that its decay width lies within the interval from 40 MeV to 100
MeV.

The decay width of $\sigma_{III}$ is slightly
different for both cases.
In case $\sigma_{III}$ is $f_0(1500)$, we have
\beq
\Gamma_{\sigma_{III}\to\pi\pi}=96\; \mbox{MeV}, \quad \Gamma_{\sigma_{III}\to KK}=176\;
\mbox{MeV},
\eeq
and in the other case ($\sigma_{III}\equiv f_0(1710)$)
\beq
\Gamma_{\sigma_{III}\to\pi\pi}=120\; \mbox{MeV}, \quad \Gamma_{\sigma_{III}\to KK}=160\;
\mbox{MeV}.
\eeq

As one can see, we obtained reasonable values for the states which are mostly
quarkonia and overestimated decay width for the state $\sigma_{III}$.
This can be explained by that the mixing of $s\bar s$ quarkonium with the
glueball
 in this type of model is too large (it is proportional to the cubed mass of
strange quark, see (\ref{BAchi})).
As a result, the decay $\sigma_{III}\to KK$ is large.
This mixing becomes a bit
less, when we fit the paramers for a higher glueball mass.

Let us note also that we do not include the decay into $4\pi$. This process
is not dominant in our model (contrary to Ref.~\cite{Shakin}).
Our estimates are based on the assumption that the process $\sigma_{III}\to 4\pi$
occurs through two intermediate $\sigma$ resonances $\sigma_{III}\to\sigma\sigma\to
4\pi$. We found that $\Gamma_{\sigma_{III}\to 4\pi}$ does not exceed 20 MeV.
Therefore, in both the cases, the total width of $\sigma_{III}$ is approximately
$300$ MeV.

\section{Conclusion}
In this work, we investigated a possible way of including
the glueball into an effective chirally symmetric meson Lagrangian.
This Lagrangian was studied in Ref.~\cite{Cimen_99} where
the masses and strong decay widths of the ground scalar meson states
were estimated in the NJL model with the 't Hooft interaction taken into account.
Now, following Ref.~\cite{Cimen_99}, we considered the interaction of
the glueball with the ground scalar isoscalar
$q\bar q$ states  $f_0(400-1200)$ and $f_0(980)$.
The mixing of the glueball with
radially excited states $f_0(1370)$ and $f_0(1710)$ (if $\sigma_{III}\equiv f_0(1500)$)
was not taken into account. However, their mixing is very
important and will be considered in  subsequent works. Therefore,
the results obtained here are tentative and are not claimed for
a quantitative explanation of experimental data on scalar
resonances.

As it was mentioned in the Introduction, nowadays there are
many papers devoted to the description of
the scalar glueball in the framework of an effective meson Lagrangian
\cite{Kusa_93,Andr_86,Jami_96,Elli_84}.
The way, we introduce the scalar glueball, is closer to that used in
Ref.~\cite{Jami_96}, but our work
differs  in two points. First, we  take into
account the 't Hooft interaction leading to the singlet-octet
mixing of the scalar isoscalar quarkonia. The glueball is
also involved in this mixing and changes it. Next, we introduce
the glueball as a dilaton field into the Lagrangian written
in a chirally symmetric form corresponding to the phase with
chiral symmetry not broken spontaneously. Thereby, the Lagrangian
is given  a highly symmetric form that keeps  both
the chiral symmetry and scale invariance. In
this form the dilaton fields are introduced only into
the mass terms of meson Lagrangian. The rest of Lagrangian terms
are scale invariant except the term with the current quark mass which
explicitly breaks both the chiral symmetry and scale invariance.

As a result, we obtain reasonable estimates for the masses of the
scalar mesons $f_0(400-1200)$, $f_0(980)$%
\footnote{In Ref.~\cite{Jami_96}, the ground $q\bar q$ scalar states
were presented by $f_0(980)$ and $f_0(1300)$ as $u\bar u$ and $s\bar s$
quarkonia mixed with the glueball.},
and the glueball $f_0(1500)$ (or $f_0(1710)$)
and also for their strong decay widths. But we have to point out
that the width of
the $f_0(1500)$ (or $f_0(1710)$) resonance is,
possibly, overestimated ($\sim 300$ MeV),
however, it can change after the states $f_0(1370)$ and $f_0(1710)$ (or $f_0(1500)$)
are included into the whole picture.

The results that we obtained in this work, as one can see,
are not enough to answer the question: which of the states
$f_0(1500)$ and $f_0(1710)$ is the scalar  glueball? We hope
to make closer towards the solution of this problem in our further works,
where alternative ways of including the glueball into
an effective meson Lagrangian will be investigated,
radially excited meson states will be considered as well as the
ground states, and the mixing of five
scalar isoscalar states $f_0(400-1200)$,
$f_0(980)$, $f_0(1370)$, $f_0(1500)$, and $f_0(1710)$
will be taken into account.

\section*{Acknowledgement}
We are grateful to Prof.~S.B.~Gerasimov and Dr.~A.Dorokhov for useful discussions. The work is
supported by RFBR Grant 00-02-17190, Heisenberg-Landau program 2000 and
by the Slovak Grant Agency for Science, Grant 2/7157/20.


\begin{thebibliography}{99}
\bibitem{Sexton} J. Sexton, A. Vaccarino, D. Weingarten,
Phys.~Rev.~Lett.~75 (1995) 4563.
%
\bibitem{LeeWeingarten} W. Lee and D. Weingarten, Phys.~Rev.~D 59 (1999)
094508.
%
\bibitem{VaccarinoWeingarten} A. Vaccarino and D. Weingarten, hep-lat/9910007.
%
\bibitem{Amsler} C. Amsler et~al. Phys.~Lett.~B 353 (1995) 425;
C. Amsler and F. Close, Phys.~Lett.~B 353 (1995) 385.
%
\bibitem{Naris_98} S. Narison, Nucl.~Phys.~B 509 (1998) 312.
%
\bibitem{Aniso_98} V. V. Anisovich, D. V. Bugg and A. V. Sarantsev,
Phys.~Rev.~D 58 (1998) 111503.
%
\bibitem{Palano} A. Palano, Nucl.~Phys.~Proc.~Suppl.~39BC (1995) 287.
%
\bibitem{Tornqvist} N.A.~T\"ornqvist, M. Roos, Phys.~Rev.~Lett.~76 (1996) 1575.
%
\bibitem{Beveren} Eef van Beveren and G. Rupp, hep-ph/9806246.
%
\bibitem{Shakin} L. S. Celenza, B. Huang, H. Wang, and C. M. Shakin,
preprint BCCNT:99/111/283, Brooklin Coll., N.Y., 1999.
%
\bibitem{Volk_991} M. K. Volkov and V. L. Yudichev, Excited scalar mesons
in a chiral quark model,
Int.~J.~Mod.~Phys.~A 14 (1999) 4621.
%
\bibitem{Volk_992} M. K. Volkov and V. L. Yudichev, Radial excitations
of scalar and $\eta$, $\eta'$ mesons in a chiral quark model,
Phys.~At.~Nucl.
v.~63(8), 2000 (at press); hep-ph/9905368.
%
\bibitem{Volk_993} M. K. Volkov and V. L. Yudichev: Radially excited
scalar, pseudoscalar and vector meson nonets in a chiral quark model.
Phys.~Part.~Nucl., v.~31(3), 2000  (at press); hep-ph/9906371.
%
\bibitem{Volk_82} M. K. Volkov and D. Ebert, Sov.~J.~Nucl.~Phys. 36
(1982) 736; Z. Phys. C16 (1983) 205;
M. K. Volkov, Ann. Phys. 157 (1984) 282.
%
\bibitem{Volk_86} M. K. Volkov, Sov. J. Part. and Nuclei 17 (1986) 186;
D. Ebert and H. Reinhardt, Nucl.~Phys.~B 271 (1986) 188;
D. Ebert, H. Reinhardt, M.K. Volkov, Prog.~Part.~Nucl.~Phys.~33 (1994) 1.
%
\bibitem{Weiss} M. K. Volkov, C. Weiss, Phys.~Rev.~D56 (1997) 221.
%
\bibitem{YaF} M. K. Volkov, Phys.~At.~Nucl.~60 (1997) 1920.
%
\bibitem{Kusa_93} K. Kusaka, M. K. Volkov and W. Weise, Phys.~Lett.
B 302 (1993) 145.
%
\bibitem{Andr_86} A. A. Andrianov, V. A. Andrianov, V. Yu. Novozhilov
and Yu. V. Novozhilov, JETP Lett. 43 (1986) 719;
A. A. Andrianov, V. A. Andrianov, D. Ebert and T. Feldmann, Int.~J.~Mod.~Phys.~A
12 (1997) 5589;
A. A. Andrianov and V. A. Andrianov, Z. Phys.~C 55 (1992) 435;
A. A. Andrianov, V. A. Andrianov, Yu.~V. Novozhilov and V. Yu.
Novozhilov, Phys.~Lett.~B 186 (1987) 401.
%
\bibitem{Jami_96} J. Cugnon, M. Jaminon and B. Van den Bosche,
Nucl.~Phys.~A 598 (1996) 515;
M. Jaminon and B. Van den Bosche, Nucl.~Phys.~A 619 (1997) 285.
%
\bibitem{Elli_84} J. Ellis and J. L\'anik, Phys.~Lett.~B 150 (1984)
289;
J. Ellis and J. L\'anik, Phys.~Lett.~B 175 (1986) 83;
J. L\'anik, Acta Phys.~Slov.~35 (1985) 343; JINR Rapid Comm.
N20-86 (1986) 10.
%
\bibitem{Cimen_99} M. K. Volkov, M. Nagy and V. L. Yudichev,
Nuovo Cim.~A 112 (1999) 225.
%
\bibitem{OPE}   M. A. Shifman, A. I. Vainshtein and V. I. Zakharov,
Nucl.~Phys.~B 120 (1977) 316.
%
\bibitem{Vogl_91} H.~Vogl and W. Weise, Progr.~Part.~Nucl.~Phys.
27 (1991) 195.
%
\bibitem{Kleva_92} S. P. Klevansky, Rev.~Mod.~Phys.  64
(1992) 649.
\bibitem{Colli_77} J. Collins, A. Duncan and S. D. Joglekar, Phys.~Rev.~D
16 (1977) 438;
N. K. Nielsen, Nucl.~Phys. B 120 (1977) 212.
%
\bibitem{Volk_93} M. K. Volkov, Phys.~Part.~Nucl. 24 (1993) 35.
%
\bibitem{Narison96} D. J. Broadhurst et al., Phys.~Lett.~B 329 (1994) 103;
B. V. Geshkenbein, Phys.~At.~Nucl.~58 (1995) 1171; S. Narison, Phys.~Lett.~B
387 (1996) 162.

\end{thebibliography}
\end{document}